\documentclass[10pt,twocolumn,showpacs,aps,prl,
               reprint,
               superscriptaddress,raggedbottom,floatfix,longbibliography]{revtex4-1}
\usepackage[T1]{fontenc}
\usepackage[utf8]{inputenc}
\usepackage{graphicx}
\usepackage{bm}

\usepackage{hyperref}
\usepackage{epstopdf}
\usepackage[version=4]{mhchem}
\usepackage{siunitx}
\usepackage{multibib}
\newcites{S}{Supplementary References}  

\usepackage{my_acronyms}

\usepackage[utf8]{inputenc}

\graphicspath{{./}{figures/}}

\newcommand{\I}{\mathrm{i}}       

\newcommand{\ket}[1]{\mathinner{|{#1}\rangle}}
\newcommand{\op}[1]{{}\widehat{\bm{#1}}}
\newcommand{\mat}[1]{\bm{#1}}
\newcommand{\vect}[1]{\vec{\bm{#1}}}

\newcommand{\pdiff}[3][{}]{\frac{\partial^{#1}{#2}}{\partial{#3}^{#1}}}
 
\newcommand{\ua}{\uparrow}    
\newcommand{\ub}{\downarrow}  

\begin{document}
\title{Negative-Mass Hydrodynamics in a Spin-Orbit--Coupled Bose-Einstein Condensate}

\author{M.~A.~Khamehchi}
\affiliation{Department of Physics and Astronomy, Washington State University,%
  Pullman,  WA 99164, USA}

\author{Khalid~Hossain}
\affiliation{Department of Physics and Astronomy, Washington State University,%
  Pullman,  WA 99164, USA}

\author{M.~E.~Mossman}
\affiliation{Department of Physics and Astronomy, Washington State University,%
  Pullman,  WA 99164, USA}

\author{Yongping~Zhang}
\email{yongping11@t.shu.edu.cn}
\affiliation{Quantum Systems Unit, OIST Graduate University,%
  Onna, Okinawa 904-0495, Japan}
\affiliation{Department of Physics, Shanghai University,%
  Shanghai 200444, China}

\author{Th.~Busch}
\email{thomas.busch@oist.jp}
\affiliation{Quantum Systems Unit, OIST Graduate University,%
  Onna, Okinawa 904-0495, Japan}

\author{Michael~McNeil~Forbes}
\email{michael.forbes@wsu.edu}
\affiliation{Department of Physics and Astronomy, Washington State University,%
  Pullman,  WA 99164, USA}
\affiliation{Department of Physics, University of Washington,%
  Seattle,  WA 98105, USA}

\author{P.~Engels}
\email{engels@wsu.edu}
\affiliation{Department of Physics and Astronomy, Washington State University,%
  Pullman,  WA 99164, USA}

\begin{abstract}
  A negative effective mass can be realized in quantum systems by engineering the dispersion relation.
  A powerful method is provided by spin-orbit coupling, which is currently at the center of intense research efforts.
  Here we measure an expanding spin-orbit coupled Bose-Einstein condensate whose dispersion features a region of negative effective mass.
  We observe a range of dynamical phenomena, including the breaking of parity and of Galilean covariance, dynamical instabilities, and self-trapping.
  The experimental findings are reproduced by a single-band Gross-Pitaevskii simulation, demonstrating that the emerging features -- shockwaves, soliton trains, self-trapping, etc.\@ -- originate from a modified dispersion.
  Our work also sheds new light on related phenomena in optical lattices, where the underlying periodic structure often complicates their interpretation.
\end{abstract}

\maketitle
\noindent
Newton's laws dictate that objects accelerate in proportion to the applied force.
An object's mass is generally positive, and the acceleration is thus in the same direction as the force.
In some systems, however, one finds that objects can accelerate \emph{against} the applied force, realizing a \emph{negative} effective mass related to a negative curvature of the underlying dispersion relation.
Dispersions with negative curvature are playing an increasingly important role in quantum hydrodynamics, fluid dynamics, and optics~\cite{Wyatt:1999, Lowman:2013, Conforti:2013, Conforti:2014, Conforti:2014a, Malaguti:2014, Conforti:2015, El:2016a}.
Superfluid \glspl{BEC} provide a particularly lucrative playground to investigate this effect, due to their high reproducibility, tunability, and parametric control.
In this Letter we report on the experimental observation of negative-mass dynamics in a \gls{SOC} \gls{BEC}.
Modeling the experiments with a single-band \gls{GPE}, we clarify the underlying role of the dispersion relation.

\begin{figure}[tb]
  \includegraphics*[width=\columnwidth]{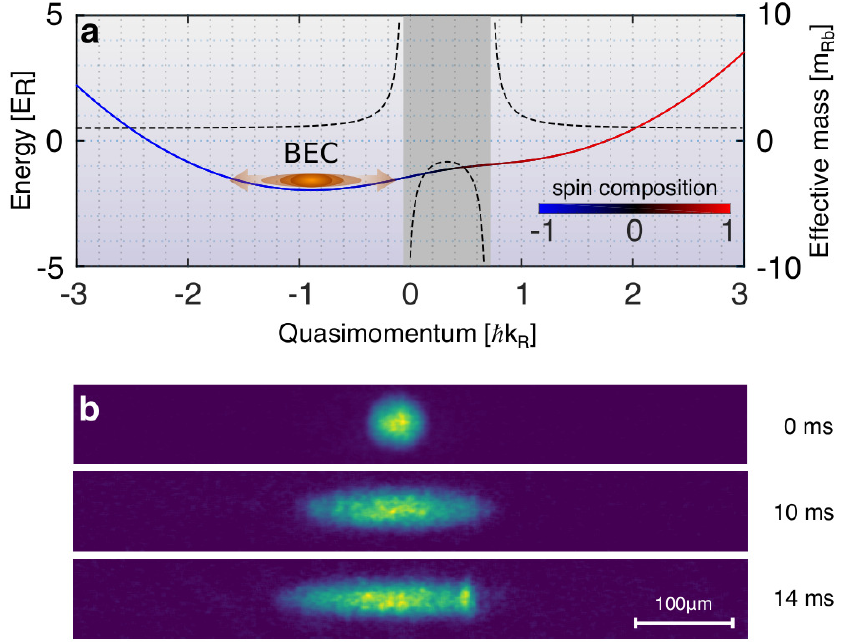}
  \caption{\label{fig:introfig}%
    (a)~Schematic representation of the 1D expansion of a \gls{SOC} \gls{BEC}. 
    The asymmetry of the dispersion relation (solid curve) causes an asymmetric expansion of the condensate due to the variation of the effective mass. 
    The dashed lines indicate the effective mass, and the shaded area indicates the region of negative effective mass. 
    The parameters used for calculating the dispersion are $\Omega = 2.5E_R$ and $\delta = 1.36E_R$. 
    The color gradient in the dispersion shows the spin polarization of the state. 
    (b)~Experimental \gls{ToF} images of the effectively 1D expanding \gls{SOC} \gls{BEC} for expansion times of \SIlist{0;10;14}{ms}.}
\end{figure} 

\begin{figure*}[tb]
  \includegraphics*[width=\textwidth]{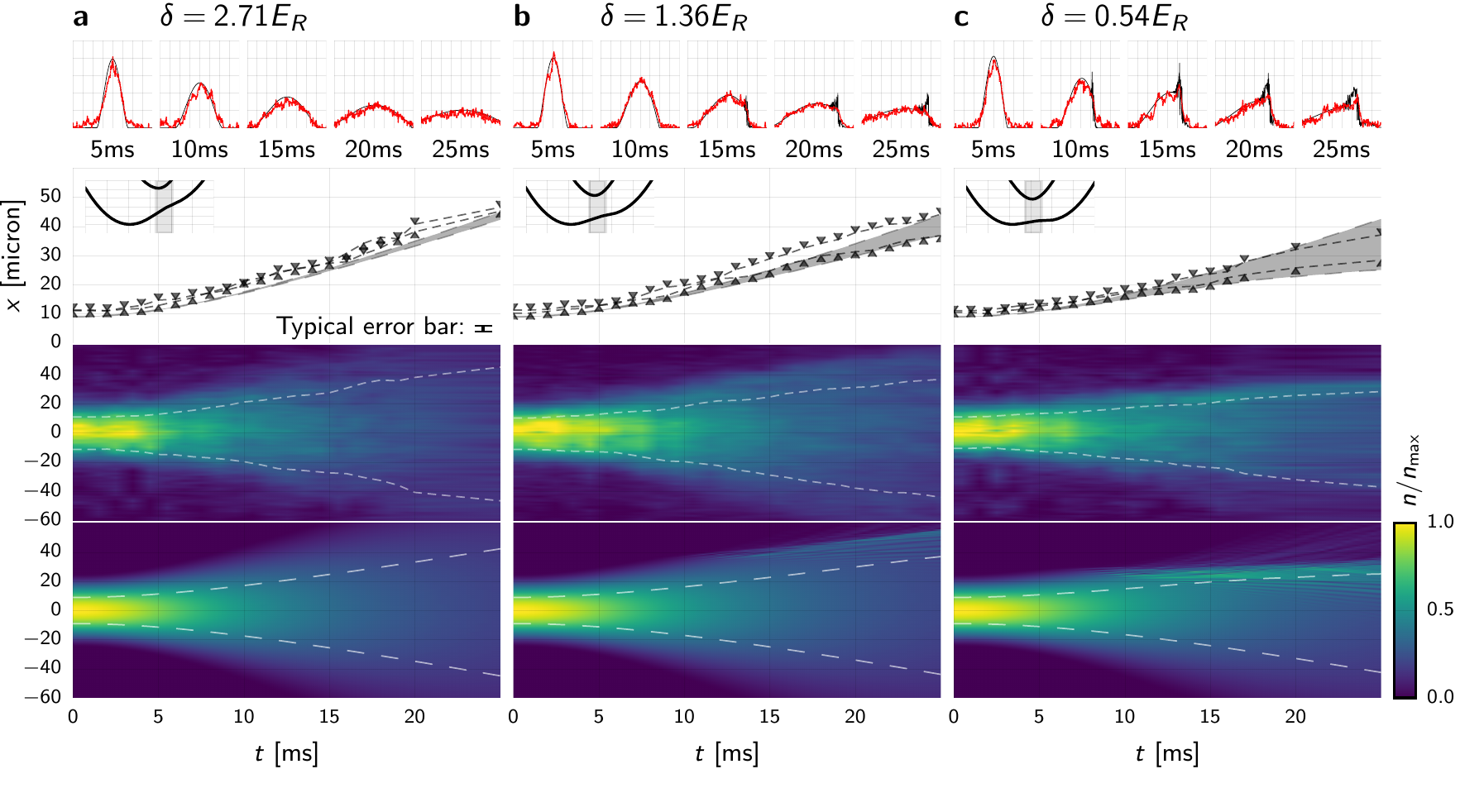}
  \caption{\label{fig:edges}%
    Anisotropic expansion of a \gls{BEC} along the direction of spin-orbit coupling ($x$ axis) with the Raman coupling strength $\Omega=2.5E_{R}$, and various detunings of %
    (a) $\delta = 2.71E_R$, %
    (b) $\delta = 1.36E_R$, and %
    (c) $\delta = 0.54E_R$, from left to right respectively.  
    The first row shows the experimental integrated cross section of the condensate (dotted red curves) overlaid with results from the \gls{GPE} simulation (solid black curves).
    The second row plots the location of the 20\% and 80\% quantiles with respect to the initial cloud center.
    The shaded regions present the \gls{GPE} simulations, while the data points (and dotted lines to guide the eye) are the experimental measurements smoothed slightly by averaging the nearest 3 times.
     Error estimates are the standard deviation of 5 samples and are comparable with about twice the camera pixel resolution.
    The insets show the dispersion relation, with the region of negative effective mass lightly shaded.
    The bottom two rows show the evolution of the expanding condensate from the experiment (upper) and corresponding single-band axially symmetric 3D \gls{GPE} simulations (lower).
    The dashed white lines depict the quantiles from the plot above.
    All experimental data presented here are from \textit{in situ} imaging.}
\end{figure*} 

We engineer a dispersion by exploiting Raman dressing techniques that lead to a \gls{SOC} \gls{BEC}~\cite{Lin:2011, Wang:2012a, Cheuk:2012, Zhang:2012, Qu:2013, Olson:2014, Hamner:2015, Luo:2015, *Luo:2016, Huang:2016, Wu:2015}.
The presence of the Raman coupling, which acts as an effective perpendicular Zeeman field, opens a gap at the crossing of two separated parabolic free-particle dispersion curves.
For suitable parameters, the lower dispersion branch acquires a double-well structure as a function of quasimomentum with a region of negative curvature [see Fig.~\ref{fig:introfig}(a)].
In our experiments, the \gls{BEC} is initially spatially well confined, then allowed to expand in one dimension in the presence of 1D spin-orbit coupling.
We observe exceptionally rich dynamics, including the breaking of Galilean covariance, which directly manifests as an anisotropic expansion of the symmetric initial state [see Fig.~\ref{fig:introfig}(b)].
As their quasimomentum increases, atoms on one side enter the negative-mass regime and slow down, demonstrating a self-trapping effect.
Related to the onset of the negative effective mass are a limiting group velocity and a dynamical instability~\cite{Fallani:2004, Bloch:2005, Bhat:2015} that leads to the formation of shockwaves and solitons.

Our results suggest that a modified dispersion and the corresponding negative effective mass also underlie various ``self-trapping'' effects seen in optical lattice experiments~\cite{Eiermann:2003, Anker:2005, Henderson:2006, Reinhard:2013, Ronzheimer:2013}.
However, in optical lattices, the origin of the self-trapping effect has been the subject of some controversy due to the occurrence of the underlying periodic potential.
This complication is absent on our work.

We start with a \gls{BEC} of approximately $10^5$ $^{87}$Rb atoms confined in a cigar-shaped trap oriented along the $x$ axis of a far-detuned crossed dipole trap (see~\cite{EPAPS} for details).
A spin-orbit coupling is induced along the $x$ axis by two Raman laser beams that coherently couple atoms in the $\ket{F,m_F} = \ket{1,-1}$ and $\ket{1,0}$ states.
A quadratic Zeeman shift effectively decouples the $\ket{1,+1}$ state from the Raman dressing, resulting in a system with pseudo-spin $\frac{1}{2}$ where we identify $\ket{\ua} = \ket{1, -1}$ and $\ket{\ub} = \ket{1, 0}$.
In this system, energy and momentum are characterized in units of $E_R = \hbar^2 k_{R}^2/{2m} \approx 2\pi\hbar \times \SI{1843}{Hz}$ and $k_{R} = 2\pi/\sqrt{2}\lambda_{R}$, respectively, where $\lambda_R = \SI{789.1}{nm}$.

Using an adiabatic loading procedure, the \gls{BEC} is initially prepared such that it occupies the lowest minimum of the lower \gls{SOC} band shown schematically in Fig.~\ref{fig:introfig}(a).
By suddenly switching off one of the two dipole trap beams, the condensate is allowed to spread out along the $x$ axis.
After various expansion times, the \gls{BEC} is imaged \textit{in situ} (see Fig.~\ref{fig:edges} and~\cite{EPAPS}), or after \SI{13}{ms} of free expansion without a trapping potential or \gls{SOC} [see Fig.~\ref{fig:introfig}(b)].
In the negative $x$ direction, the \gls{BEC} encounters an essentially parabolic dispersion, while in the positive $x$ direction, it enters a negative-mass region.
This leads to a marked asymmetry in the expansion.

Experimental results together with matching numerical simulations are presented in Fig.~\ref{fig:edges}, which shows three sets of data with Raman coupling strength $\Omega = 2.5E_R$ and Raman detunings $\delta \in \{2.71E_R, 1.36E_R, 0.54E_R\}$.  
With decreasing $\delta$, the dispersion relation (shown in the inset in the second row) develops a more pronounced double-well structure in the lower band.
The \gls{BEC} is initially placed at the global minimum of the lower band, and the expansion dynamics are initiated at time $t = 0$.
Integrated cross sections are shown in the first row, the location of the $20\%$ and $80\%$ quantiles in second row, and the experimental and numerical time-slice plots in the third and fourth rows respectively.
For $\delta = 2.71E_R$, the expansion is almost symmetric.
For $\delta = 1.36E_R$, one sees a noticeable slowing down of the positive edge of the cloud.
In the corresponding \gls{GPE} simulations this occurs at about \SI{11}{ms} when this edge of the cloud has expanded to $\SI{40}{\micro m}$ and the quasimomenta enter the negative-mass region.
Here one sees a pileup of the density and a dynamic instability, i.e. an exponential growth in the amplitude of phonon modes.
For the smallest detuning $\delta = 0.54E_R$, the effect is even more pronounced, and in the corresponding \gls{GPE} simulations, the pileup and instability start sooner at about \SI{8}{ms} and \SI{25}{\micro m}.

At the lowest Raman detuning ($\delta = 0.54E_R$), the experiments show a slowdown at approximately \SI{15}{ms} in the negative edge, which is not seen in the \gls{GPE} simulations.
We speculate that this is due to finite temperature allowing a population in the second local minimum, requiring more sophisticated simulations (see e.g.~\cite{Minguzzi:2004, *FINESS:2013}).
Aside from this effect, we see good agreement between the experiment and zero-temperature \gls{GPE} simulations, 
as is expected due to the sizable roton gap (hence small noncondensed fraction~\cite{Zhang:2016}) and weak interactions~\footnote{The scattering lengths are $a_{\ua\ua} = 100.40a_0$, $a_{\ub\ub} = 100.86a_0$, and $a_{\ua\ub} = 100.41a_0$~\cite{Kokkelmans:pc, *Verhaar:2009}, so that the dimensionless gas parameter $na^3 < \num{3e-5} \ll 10^{-3}$.}.%
\marginpar{\hspace{-100em}\cite{Kokkelmans:pc, *Verhaar:2009}}

This justifies the description of the dynamics with a coupled set of \glspl{GPE} describing the two spin components:
\begin{subequations}\label{eq:H2}
  \begin{gather}
    \I\hbar\pdiff{}{t}
    \begin{pmatrix}
      \ket{\ua}\\
      \ket{\ub}
    \end{pmatrix}
 =
   \begin{pmatrix}
     \frac{\op{p}^2}{2m} + V_\ua & \frac{\Omega}{2}e^{2\I k_R x}\\
     \frac{\Omega}{2}e^{-2\I k_R x} & \frac{\op{p}^2}{2m} + V_\ub
   \end{pmatrix}\cdot
   \begin{pmatrix}
     \ket{\ua}\\
     \ket{\ub}
   \end{pmatrix},\\
    V_{\ua/\ub} = -\mu \pm \frac{\delta}{2} + g_{\ua\ua/\ua\ub}n_\ua + g_{\ua\ub/\ub\ub}n_\ub
  \end{gather}
\end{subequations}
where $\op{p} = -\I\hbar\vect{\nabla}$ is the momentum operator, $\mu$ is a common chemical potential, $k_R$ is the Raman wave vector, $g_{ab} = 4\pi\hbar^2 a_{ab}/m$, and $a_{ab}$ are the $S$-wave scattering lengths.
For the $\ket{\ua} = \ket{1, -1}$ and $\ket{\ub} = \ket{1, 0}$ hyperfine states of \ce{^{87}Rb}, the scattering lengths are almost equal and for our numerics we take $a_{\ua\ua} = a_{\ua\ub} = a_{\ub\ub} = a_s$.
We compare our experimental results with 3D axially symmetric simulations for $\omega_\perp = 2\pi\times\SI{162}{Hz}$ and realistic experimental parameters.

One of the theoretical results we wish to convey is that in many cases (e.g., Refs.~\cite{Wang:2006, Li:2015}), a single-band model can capture the essential dynamics with modified dispersion:
\begin{gather}
  \label{eq:single-band}
  \I\hbar\pdiff{}{t} \ket{\psi} = \left[
    E_{-}(\op{p}) + gn + V_{\text{ext}}(x)
  \right]\ket{\psi}
\end{gather}
where $E_{-}(\op{p})$ is the dispersion of the lower band obtained by diagonalizing Eq.~\eqref{eq:H2} for homogeneous states.
For inhomogeneous densities this picture is locally valid for slowly varying densities, similar to the Thomas-Fermi approximation, and remains valid as long as the system is gently excited compared to the band separation, which is proportional to the strength $\Omega$ of the Raman coupling.
With our parameters, the single-band model exhibits almost identical results to the multiband description, quantitatively reproducing many aspects of the experiment.
The approximate equality of the coupling constants allows one to define a spin-quasimomentum mapping that relates the two-component spin populations $n_\ua$ and $n_\ub$ to the quasimomentum $q$ of the single-component state: 
\begin{gather}
  \label{eq:spin-quasimomentum-map}
  \frac{n_\ub - n_\ua}{n_\ub + n_\ua} = \frac{k - d}{\sqrt{{(k - d)}^2 + w^2}},
\end{gather}
where we have defined the dimensionless parameters $k = p/\hbar k_R$, $d = \delta/4E_R$, and $w = \Omega/4E_R$.
Our simulations solve this model with a \gls{DVR} basis (see, e.g., Ref.~\cite{LC:2002}) with $4096\times 128$ lattice points in a periodic tube of length \SI{276}{\micro\m} and radius \SI{10.8}{\micro\m}.
This greatly simplified analysis shows that all of the interesting phenomena observed in the experiment -- asymmetric expansion, the pileup, slowing down, and instabilities -- follow from the modified dispersion relationship.

\begin{figure}[tb]
  \includegraphics*[width=\columnwidth]{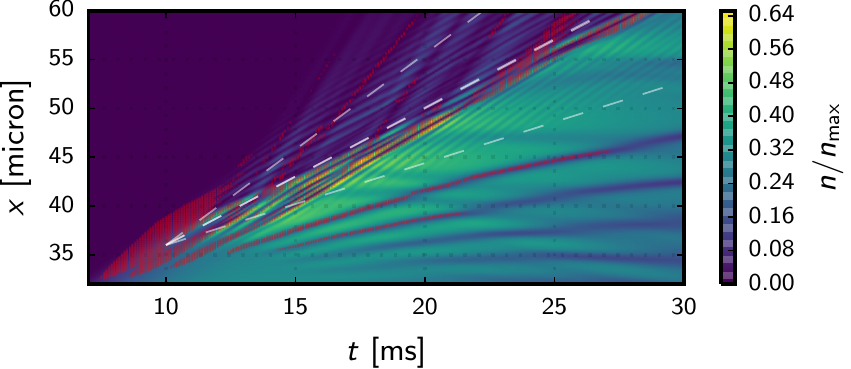}
  \caption{\label{fig:simulation1_zoom}%
    Zoom of a 1D simulation matching the lower middle frame of Fig.~\ref{fig:edges} showing the total density $n = n_\ua + n_\ub$ as a function of time in the region where the dynamic instability first appears.
    Dashed lines are the three group velocities $v_g = E_{-}'(k)$ at the quasimomenta $k$ where the inverse effective mass $m_*^{-1} = E_{-}''(k)$ first becomes negative (steepest line), the point of maximum $m_*^{-1}$ (middle), and the point where the $m_*^{-1}$ returns to positive (least steep line).
    Red points demonstrate where the local quasimomentum lies in the negative-mass region.
    Note that, as described in the text, the pileup initially contains many density fluctuations, but sharpens as solitons and phonons ``radiate'' energy away from the wall.
  }
\end{figure}

Having introduced the single-component theory, we now derive the hydrodynamics of this model by effecting a Madelung transformation $\psi = \sqrt{n}e^{\I\phi}$ where $n(\vect{x},t) = n_{\ua} + n_{\ub}$ is the total density at position $\vect{x}$ and time $t$, and the phase $\phi(\vect{x}, t)$ acts as a quasimomentum potential $\vect{p} = \hbar\vect{\nabla}\phi$.
The hydrodynamic equations are
\begin{subequations}
  \begin{gather}
    \pdiff{}{t} n + \vect{\nabla}\cdot(n\vect{v}) = 0,\\
    \pdiff{\vect{v}_*}{t}  + (\vect{v}_* \cdot \vect{\nabla})\vect{v}_* = \mat{M}_*^{-1}\cdot(\overbrace{-\vect{\nabla}[V_{\text{eff}} + V_{Q}]}^{\vect{F}}),\\
    [\mat{M}^{-1}_*]_{ij} = \frac{\partial{E_{-}}(\vect{p})}{\partial{p_i}\partial{p_j}}, \qquad
    [\vect{v}_*]_{i} = \pdiff{E_{-}(\vect{p})}{p_i},
  \end{gather}
\end{subequations}
where $\vect{v}$ is the group velocity and $\vect{j} = \vect{v} n$ is the current density.
$V_{\text{eff}} = V_{\text{ext}}(\vect{x}, t) + gn(\vect{x}, t)$ is the effective potential, including both the external potential and mean-field effects. 
What differs from the usual Madelung equations is that third and higher derivatives of the dispersion $E_{-}(\vect{p})$ affect the velocity $\vect{v}_*$ and quantum potential $V_Q(n, \vect{p})$.
While for homogeneous matter, $\vect{v} = \vect{v}_*$, this relationship is broken in inhomogeneous matter and the quantum potential acquires terms beyond the usual quantum pressure term $V_Q(n) \propto \nabla^2\sqrt{n}/\sqrt{n}$.
For approximately homogeneous sections of the cloud, however, these corrections are small: $\vect{v}\approx\vect{v}_*$ and the usual hydrodynamic behavior is realized.
In particular, $\partial\vect{v}/\partial{t} \approx \mat{M}_*^{-1}\cdot\vect{F}$, so that the group velocity responds classically, accelerating against the force $\vect{F}$ if the effective mass is negative.
The experiment may be qualitatively explained using a Thomas-Fermi--like approximation where each point of the cloud is locally described by a plane-wave with local quasimomentum $p$.
Initially, equilibrium is established between the external trapping force $-\nabla V_{\text{ext}}$ and the internal mean-field pressure $-\nabla (gn)$, with the quasimomentum $p = p_0$ minimizing the kinetic energy $E_{-}'(p_0) = 0$.
About this minimum, the effective mass is positive, so once the trapping potential $V_{\text{ext}}$ is reduced, the cloud starts to expand due to the mean-field pressure, generating an outward group velocity and quasimomenta.
As the quasimomentum along the positive $x$-axis approaches the negative-mass region (see Fig.~\ref{fig:simulation1_zoom}), the acceleration slows significantly compared with the acceleration along the negative axis, leading to the asymmetric expansion seen in Fig.~\ref{fig:edges}: a manifestation of the broken Galilean covariance and parity in \gls{SOC} systems.
Once the quasimomentum enters the negative-mass region, the acceleration opposes the force, and the cloud experiences the ``self-trapping'' effect where the positive mean-field pressure tends to prevent further expansion.

A similar effect has been reported in optical lattices near the edge of the Brillouin zone~\cite{Eiermann:2003, Anker:2005, Chang:2014}.
These ``self-trapping'' effects in lattices have been attributed to several different phenomena.
One, based on a Josephson effect with suppressed tunneling between neighboring sites~\cite{Anker:2005, Wang:2006}, was predicted from a variational framework~\cite{Trombettoni:2001}.
Another explanation is that the sharp boundary is a ``gap soliton''~\cite{Alexander:2006}, though this explanation is disputed by~\cite{Wang:2006} on the basis that solitons should remain stable whereas the latter observe self-trapping only for a finite period of time.
Finally, self-trapping has been explained in terms of the Peierls-Nabarro energy barrier~\cite{Hennig:2016}.
In all of these cases, the self-trapping occurs where the effective mass becomes negative, but the interpretation of the self-trapping effect in optical lattices is complicated by the presence of spatial modulations in the potential.

The beauty of engineering dispersions with \gls{SOC} \glspl{BEC} is that lattice complications are removed.
The success of the single-band model in reproducing the experiment demonstrates clearly that the self-trapping results from the effective dispersion relationship.
A single-band model using the dispersion of the lowest band in an optical lattice is able to explain the previous observation of self-trapping in lattice systems, clearly demonstrating the importance of the band structure and deemphasizing the role of the underlying lattice geometry of coupled wells.
This result is confirmed by using a tight-binding approximation to map the optical lattice of Ref.~\cite{Wang:2006} to a single band model, which reproduces their results.
In our simulations, although the boundary appears to be very stable, it is ``leaky''\!\!.
This can be seen from Fig.~\ref{fig:simulation1_zoom} where the boundary maintains its shape, but permits a small number of fast moving particles to escape.
Similarly, in optical lattice systems such fast moving particles are responsible for the continued increase in the width of the cloud seen by~\cite{Wang:2006} even though the boundary remains stopped.
This may resolve the apparent discrepancy between~\cite{Wang:2006} and~\cite{Alexander:2006} as a quasistable but leaky gap soliton.

What sets the limiting velocity of the expanding edge?
In the optical lattice system of Ref.~\cite{Eiermann:2003}, the limiting velocity was observed to lie at the inflection point where the mass first starts to become negative.  
In contrast, our \gls{GPE} simulations for \gls{SOC} \gls{BEC}s clearly show this limiting velocity to lie fully inside the negative-mass region, near the point of maximum negative acceleration (maximum negative inverse effective mass) (see Fig.~\ref{fig:simulation1_zoom}).
While this qualitatively describes the limiting velocity, the full effect is somewhat subtle.
The limiting velocity ultimately depends on several factors, including the preparation of the system~\cite{Delikatney:2016}. 
It requires a quasistable boundary which is tied to the negative effective mass through a dynamical instability.
From the \gls{GPE} simulations, the picture emerges that once the cloud enters the negative-mass region, small fluctuations grow exponentially forming the sharp boundary of the cloud.
Initially these growing modes appear chaotic, but as is typical with dispersive shock-waves (see~\cite{El:2016} and references therein), energy is ``radiated'' from the boundary in the form of phonons and soliton trains that are clearly visible in Fig.~\ref{fig:simulation1_zoom}.
As energy is dissipated, the boundary appears to sharpen due to nonlinear effects.
This seems critically connected to the negative effective mass as a similar boundary with positive mass dispersion broadens~\cite{Delikatney:2016}.

In conclusion, we have studied negative-mass hydrodynamics both experimentally and theoretically in an expanding spin-orbit coupled Bose-Einstein condensate.
The experimental results are quantitatively reproduced with an effective single-band zero-temperature \gls{GPE}-like model derived from the \gls{SOC} Hamiltonian.
With this model, we see that the pileup and subsequent boundary behavior are intimately related to the presence of a negative effective mass $m_*^{-1} = E_{-}''(p)$ in the effective dispersion for the band.
From linear response theory, one finds a dynamical instability closely associated with the negative effective mass that leads to the sudden increase in density.
The boundary clearly demonstrates radiation of phonons and soliton trains, which appear to remove energy from the region, thereby allowing the boundary to stabilize.
The stability of the boundary and its final velocity depend critically on the existence of a negative effective mass.
With this work, we have also clarified the interpretation of self-trapping phenomena observed in optical lattices~\cite{Anker:2005, Wang:2006}, demonstrating that this is naturally explained by a negative effective mass.
Spin-orbit coupling provides a powerful tool for engineering the dispersion $E_-(p)$ without the additional complications of spatial modulations that appear in the context of optical lattices.

\paragraph{Acknowledgements:}
We thank Mark Hoefer for useful discussions.  This work was supported in part
by a \gls{WSU} New Faculty Seed Grant, and the Okinawa Institute of Science and
Technology Graduate University.  P.~E.\@ acknowledges funding from the \gls{NSF}
under grant No.~PHY-1607495.  Y.~Z.\@ is supported in part by the
Thousand Young Talent Program of China, and the Eastern Scholar Program of
Shanghai.

\section{References}
\bibliography{local,master}

\clearpage
\setcounter{page}{0}
\section{Numerical Simulations}\noindent
In principle, since the trap is highly elongated, one can use an effective 1D simulation by tuning the coupling constants appropriately as described in~\citeS{Olshanii:1998, Zhang:2014}.
For the experimental parameters, this effective 1D approximation works quite well, but since there are some radial excitations, we compare directly with 3D axially symmetric simulation with $\omega_\perp = 2\pi\times\SI{162}{Hz}$.
Our simulations are performed using an axially symmetric \gls{DVR} basis (see e.g.~\cite{LC:2002}) with $4096\times 128$ lattice points in a periodic tube of length \SI{276}{\micro\m} and radius \SI{10.8}{\micro\m}.

\begin{figure}[b]
  \includegraphics*[width=\columnwidth]{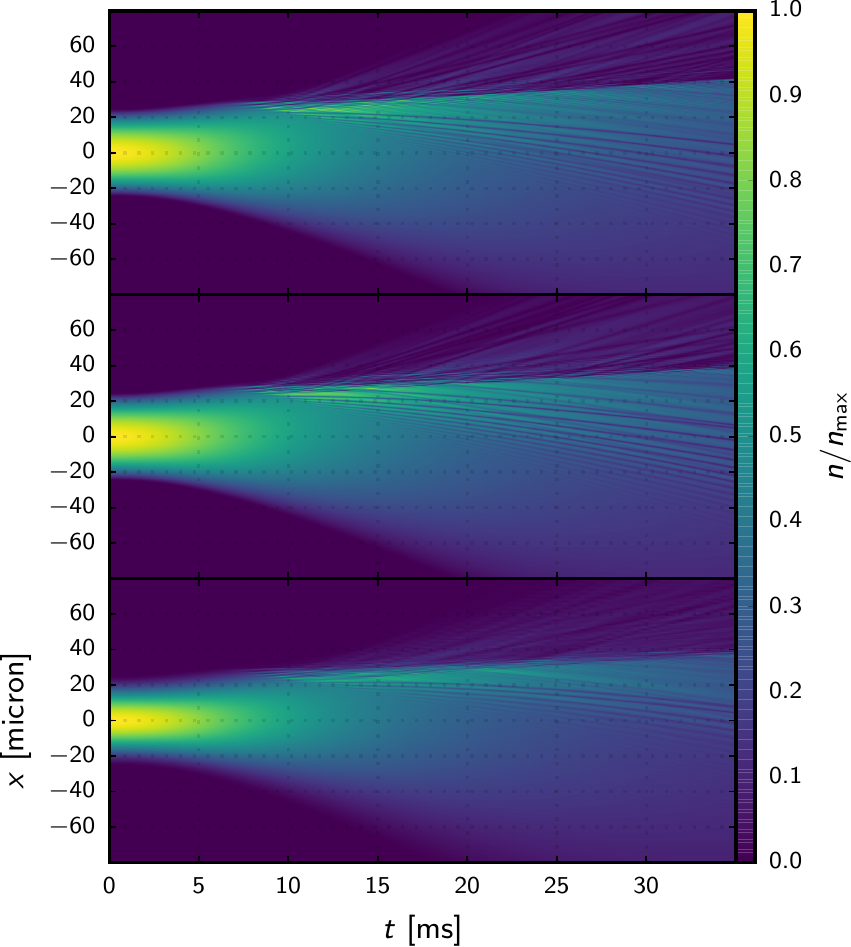}
  \caption{\label{fig:components}%
    \textbf{Comparison between simulations.}  Comparison of the full
    two-component 1D \gls{GPE} model~\eqref{eq:H2} (top) with the 1D
    single-band model~\eqref{eq:single-band} (middle) and the axially symmetric
    3D single-band model (bottom) for detuning $\delta = 0.54E_R$ and Raman 
    coupling strength $\Omega = 2.5E_R$. }
\end{figure} 

In Fig.~\ref{fig:components} we compare the single-band model~\eqref{eq:single-band} to the full two-component model~\eqref{eq:H2}, demonstrating that for the experiment under consideration, all the relevant phenomena result purely from the modified dispersion relationship $E_-(\op{p})$.
Even when studying systems with significant population of the upper band such as the dynamical spin-density waves demonstrated in~\cite{Li:2015}, we find that the single-component model~\eqref{eq:single-band} captures most of the bulk effects: i.e\@. it correctly models the dynamical behavior of the total density $n_\ua + n_\ub$, but the significant population of the upper branch breaks the spin-quasimomentum mapping~\eqref{eq:spin-quasimomentum-map}.
In this figure we also compare 3D axially-symmetric simulations with 1D simulation where we have tuned the coupling constant using the relationship in~\citeS{Olshanii:1998}.
This comparison demonstrates that, while the 1D simulations qualitatively match the experiments, they exhibit some quantitative differences with the 3D simulations.

Our simulations here explore only the dynamics of the condensate.
Extensions to the \glspl{GPE} (see e.g.~\cite{Minguzzi:2004, *FINESS:2013} for reviews) are require to capture the dynamics of the non-condensed fraction due to finite temperature, and quantum fluctuations, but are beyond the score of the present work.
(For example, a significant non-condensed fraction is expected when the gap in the roton branch vanishes~\cite{Zhang:2016}.  For our parameters, the roton branch still has a large gap.)
The deviations described in the text between \gls{GPE} simulations and experiment at the lowest Raman detuning will provide a way to test these extensions.

\begin{figure}[b]
  \includegraphics*[width=\columnwidth]{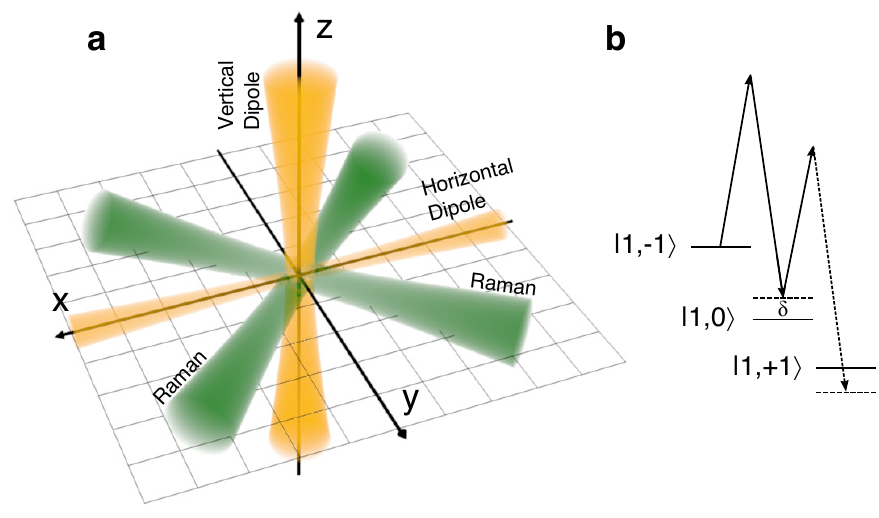}
  \caption{\label{fig:schematic}
    (a)~Experimental arrangement of the trap (yellow) and the Raman (green) beams.
    The angle between the Raman beams is $90^\circ$.
    (b)~Raman coupling scheme in the $F=1$ manifold.
    The $\ket{1,+1}$ state is effectively decoupled due to the quadratic Zeeman shift induced by an external bias field along the $x$ direction.}
\end{figure} 

\section{Experimental Methods}\noindent
The experiment begins with approximately $10^5$ $^{87}$Rb atoms confined by two trapping dipole beams which together produce harmonic trapping frequencies of $\left(\omega_x, \omega_y, \omega_z \right) = 2\pi \times$\SIlist[list-units=brackets]{26;170;154}{Hz}. 
\gls{SOC} is then generated along the $x$-axis by two perpendicular Raman beams of wavelength $\lambda_R = \SI{789.1}{nm}$ [see Fig.~\ref{fig:schematic}(a)].
The Raman beams coherently couple the  $\ket{F,~m_F} = \ket{1, -1} \text{and} \ket{1,0}$ states.
A homogeneous bias field of $B \approx \SI{10}{G}$ is applied along the $x$-axis, leading to a Zeeman splitting of the atomic levels.  
The quadratic Zeeman shift of $7.8E_R$ effectively decouples the $\ket{1,+1}$ state from the Raman dressing, resulting in a system with pseudo-spin $\frac{1}{2}$ where we identify $\ket{\ua} = \ket{1, -1}$ and $\ket{\ub} = \ket{1, 0}$ [see Fig.~\ref{fig:schematic}(b)].  
After the atoms are dressed by the Raman beams with a specific Raman coupling strength $\Omega$ and detuning $\delta$, the vertical dipole beam is rapidly switched off, reducing the confinement along the $x$-axis from $\omega_x = 2\pi \times \SI{26}{Hz}$ to $\omega_x = 2\pi \times \SI{1.4}{Hz}$.  
This allows the \gls{SOC} \gls{BEC} to expand along the direction of the spin-orbit coupling, revealing the rich dynamics discussed in the main text.
The extent of the cloud as depicted in Fig.~\ref{fig:edges} is taken from in-trap images.  Figure~\ref{fig:introfig}(b) depicts experimental images taken after an in-trap expansion time denoted to the left of the image, followed by \SI{13}{ms} free expansion.  

\begin{figure*}[tbp]
  \includegraphics*[width=\textwidth]{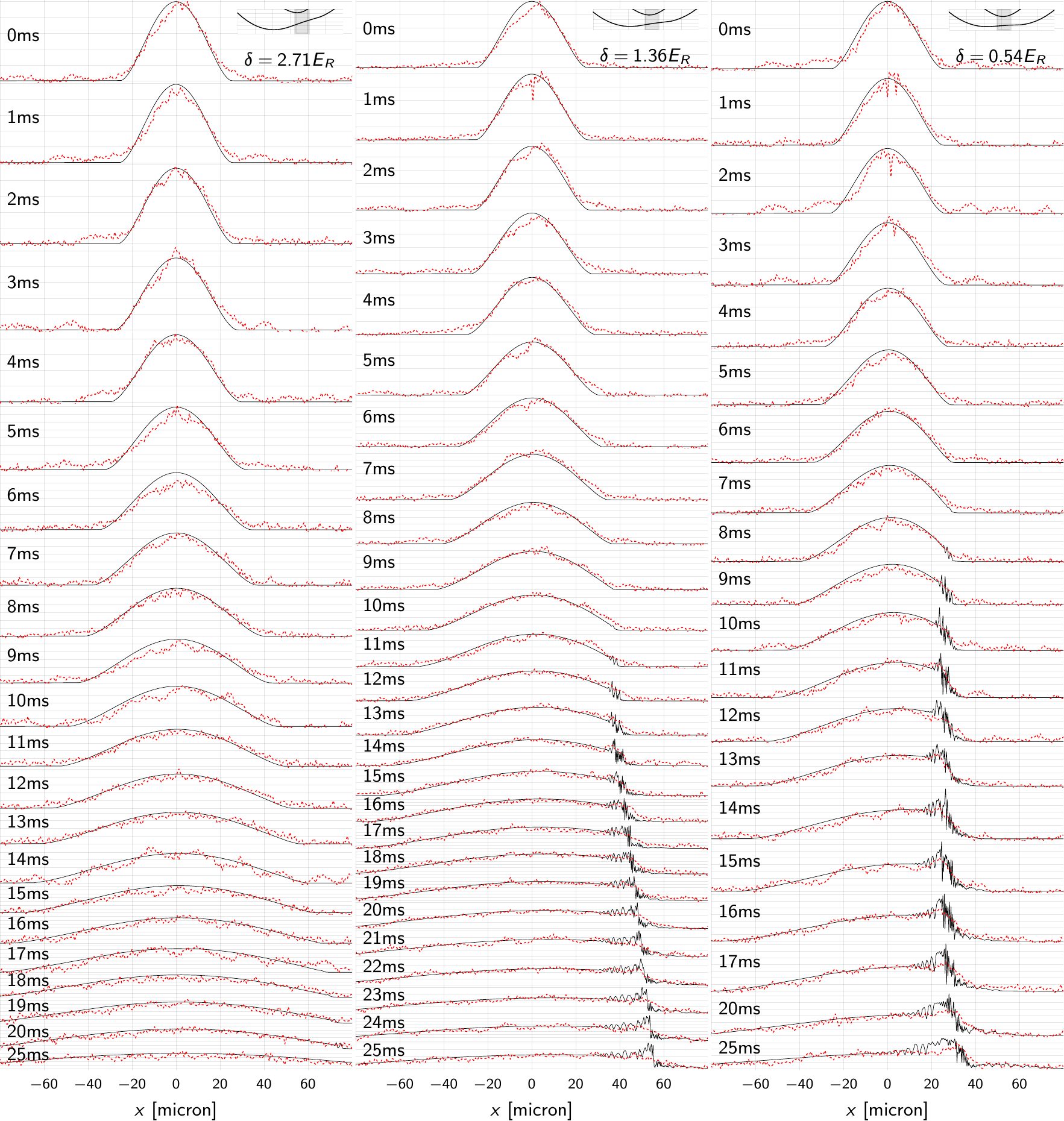}
  \caption{\label{fig:slices}%
    \textbf{In-situ data: Experiment and Theory.} 
    In-situ experimental integrated cross section of the condensate (dotted red curves) overlaid with results from the axially symmetric single-band 3D \gls{GPE} simulations (solid black curves) with the Raman coupling strength $\Omega=2.5E_{R}$, and various detunings of %
    (a) $\delta = 2.71E_R$, %
    (b) $\delta = 1.36E_R$, and %
    (c) $\delta = 0.54E_R$, from left to right respectively.
    The top insets show the dispersion relation, with the region of negative effective mass lightly shaded.
  }
\end{figure*}

\section{Comparison}\noindent
To supplement the comparison between the numerical and experimental results, we include in Fig.~\ref{fig:slices} a high-resolution comparison between the numerical and experimental \textit{in situ} images.  Note that we have not added noise or pixel averaging effects to the numerical simulations so that underlying features (solitons etc.) remain unobscured.

\section{Supplementary References}
\bibliographystyleS{apsrev}
\bibliographyS{local,master}

\end{document}